\documentclass[12pt,reqno]{amsart}
\usepackage{amsmath, amsthm, amssymb}
\usepackage{amsaddr}
\usepackage[version=3]{mhchem}
\usepackage{setspace}
\usepackage{graphics, epsfig, subfigure, epstopdf}
\usepackage{graphicx}
\usepackage{curves}
\usepackage{graphics}

\numberwithin{equation}{section}

\doublespace

\begin{document}
\author{Joel Bandstra, Ying Li and Naiyi Wu}
\address{Department of Mathematics and Engineering\\
Saint Francis University\\
Loretto, PA 15940\\}
\email{JBandstra@francis.edu, YLi@francis.edu, nxw102@francis.edu}

\title{Mathematical Modeling on Open Limestone Channel}
\begin{abstract}

Acid mine drainage (AMD) is the outflow of acidic water from metal mines or coal mines. When exposed to air and water, metal sulfides from the deposits of the mines are oxidized and produce acid, metal ions and sulfate, which lower the pH value of the water. An open limestone channel (OLC) is a passive and low cost way to neutralize AMD. The dissolution of calcium into the water increases the pH value of the solution. A differential equation model is numerically solved to predict the variation of concentration of each species in the OLC solution. The diffusion of Calcium due to iron precipitates is modeled by a linear equation. The results give the variation of pH value and the concentration of Calcium.

\end{abstract}
\maketitle

\section{Introduction}\label{intro}

Acid mine drainage (AMD) is the outflow of acidic water from metal mines or coal mines. When exposed to air and water, metal sulfides from the deposits of the mines are oxidized and produce acid, metal ions and sulfate, which lower the pH value of the water. The water is then unfit for consumption, agricultural or industrial use. AMD has become one of the largest problems for the mining industry. 

There have been various methods to treat AMD. Some of them are expensive and hard to maintain. An open limestone channel (OLC) is a passive and low cost way to treat AMD. Coarse limestone boulders are placed into downstream water. The dissolution of calcium into the water will increase the pH value of the solution. Meanwhile with time passing, the oxidization of metal causes the accumulation of metal sediment at the surface of limestone which hinders the dissolution of calcium. Studies in the 1970’s ~\cite{P-M1974}, ~\cite{P-M1975a}, ~\cite{P-M1975b} showed that limestone armored with metal oxides releases calcium at a 20\% rate of the unarmored one. Experiments conducted by Zimekiewicz et al. ~\cite{Z-S-R} investigated the factors that affect the performance of the open limestone treatment. 

We study the problem from a mathematical point of view. Numerous experiments have been done with data collected and analyzed. However no mathematical model has been utilized in these studies. The purpose of this paper is to set up a mathematical model which describes and predicts the performance of open limestone channel in treatment of AMD. Our model computes the variation of concentrations of sample species in the solution of a downstream channel. The plan of this paper is as follows. Section \ref{chemreact} explains the chemical and the mechanism of an open limestone channel. The underlying differential equation is presented in section \ref{rateeq}. Numerical results are given in section \ref{num}.

\section{Chemical Reaction}\label{chemreact}
The dissolution of limestone is the main chemical reaction occurring in open limestone channel. It can be described by the following chemical formulas ~\cite{C-T}
\begin{align}
\cee{&CaCO3_{(s)} + 2H+ <-> Ca^{2+} + H2CO3^{*}},\label{Ca1}\\
\cee{&CaCO3_{(s)} + H2CO3^{*} <-> Ca^2+ + 2HCO3-},\label{Ca2}\\
\cee{&CaCO3_{(s)} + H2O <-> Ca^2+ +HCO3- + OH-},\label{Ca3}
\end{align}
where $\ce{[H2CO3^{*}]=[CO2_{(aq)}] + [H2CO3^0]}$. Square brackets [ ] define concentrations in solution, and curly brackets \{ \} will be used to denote activities, both in mol/L. 

The metal sediment at the surface of limestone is considered to be \ce{Fe(OH)3} produced by iron ions. The reaction is as follows
\begin{align}
\cee{Fe^3+ + 3HCO3^- -> Fe(OH)3_{(s)} + 3CO2}.
\end{align}

Assuming the system is at chemical equilibrium, we can calculate the concentrations of all species in solution by applying mass balance and charge balance equations in the system. In the following we will set up a chemical equilibrium model in our system. Readers are referred to ~\cite{S-Z} for a more detailed explanation of the numerical method to solve chemical equilibrium problems.  

To compute the concentrations of species, we consider the following reactions
\begin{align}
&\ce{H2CO3^{*} <=> H+ + HCO3-},\\
&\ce{HCO3^- <=> H+ + CO3^2-},\\
&\ce{H2O <=> H+ + OH^-}, 
\end{align}
with the following equilibrium constants, respectively
\begin{align}
&K_1=\frac{\{\ce{H^+}\}\{\ce{HCO3-}\}}{\{\ce{H2CO3^{*}}\}},\\
&K_2=\frac{\{\ce{H^+}\}\{\ce{CO3^2-}\}}{\{\ce{HCO3-}\}},\\
&K_{\rm w}=\ce{\{H+\}\{OH^-\}}.
\end{align}
The species in the solution are assumed to be: \ce{[H+]}, \ce{[OH-]}, \ce{[H2CO3^{*}]}, \ce{[HCO3-]}, \ce{[CO3^2-]}, \ce{[Ca^2+]}. They satisfy the mass balance equation with respect to the concentration of \ce{Ca^2+} and the charge balance equation
\begin{align}
[\ce{H+}]+[\ce{2OH-}]=[\ce{OH-}]+[\ce{HCO3-}] + [\ce{2CO3^2-}].
\end{align}
Writing the chemical equilibirum equation for each species in $\log$ form, we have
\begin{align}
&\log [\ce{H+}]=\log [\ce{H+}]+\log \kappa_1,\\
&\log [\ce{OH-}]=-\log [\ce{H+}]+\log \kappa_2,\\
&\log [\ce{H2CO3}]=\log [\ce{H2CO3^{*}}]+\log \kappa_3,
\end{align}
\begin{align}
&\log [\ce{HCO3-}]=\log [\ce{H2CO3^{*}}]+ \log [\ce{H+}]+\log \kappa_4, \\
&\log [\ce{CO3^2-}]=\log [\ce{H2CO3^{*}}] -2\log [\ce{H+}]+\log \kappa_5, \\
&\log [\ce{Ca^2+}]=\log [\ce{Ca^2+}]+\log \kappa_6,
\end{align}
where $\kappa_i, i=1,..., 6$ are the equilibrium constants for the corresponding reactions. It is obvious that $\log \kappa_1=\log \kappa_3=\log \kappa_6 =0$. The others are defined by experiments as functions depending on temperature $T$ (K) \cite{H-S}
\begin{align}
&\log \kappa_2=\log K_{\rm w}=6.0875-\frac{4470.99}{T}-0.01706T,\\
&\log \kappa_4=14.8435-\frac{3404.71}{T}-0.03279T,\\
&\log \kappa_5=21.3415-\frac{6307.1}{T}-0.05658T.
\end{align}
Adding the two balance equations, we obtain a nonlinear system of equations with $6$ unknowns and $6$ equations. We then solve the nonlinear system of equation by Newton-Raphson method.

\section{Rate equation}\label{rateeq}
The rate of dissolution of calcite $r$ (${\rm mmol}/{\rm cm}^2/$s), determined by the chemical reactions \eqref{Ca1}, \eqref{Ca2}, \eqref{Ca3}, is given by the following equation ~\cite{P-W-P}
\begin{align}
r=k_1 a_{\ce{H+}}+k_2 a_{\ce{H2CO3^\ast}}+k_3 a_{\ce{H2O}}-k_4 a_{\ce{Ca^2+}} a_{\ce{HCO3-}},
\end{align}
where $a_{\ce{H+}}$, $a_{\ce{H2CO3^\ast}}$, $a_{\ce{H2O}}$, $a_{\ce{Ca^2+}}$, $a_{\ce{HCO3-}}$ are the activities of \ce{H+}, \ce{H2CO3^{$\ast$}}, \ce{H2O}, \ce{Ca^2+} and \ce{HCO3-}, which are assumed in this work to be the same as concentrations. The first order rate constants $k_1$, $k_2$, $k_3$ (cm/s) are defined as
\begin{align}
&\log k_1=0.198-\frac{444}{T},\\
&\log k_2=2.84-\frac{2177}{T},\\
&\log k_3=
\begin{cases}
\displaystyle -5.86-\frac{317}{T} &\quad 5\,^\circ \mathrm{C}<=T<=25\,^\circ \mathrm{C},\\ 
\displaystyle -1.10-\frac{1737}{T} &\quad 25\,^\circ \mathrm{C}<T<=48\,^\circ \mathrm{C}.\\
\end{cases}
\end{align}
The rate constant of precipitation reaction $k_4$ (cm/s) is defined as
\begin{align}
k_4=\frac{K_2}{Kc} [k'_1+\frac{1}{a_\ce{H+}}(k_2 a_{\ce{H2CO3^{$\ast$}}}+k_3 a_{\ce{H2O}})],
\end{align}
where $K_2$ is the equilibrium constant for dissociation of bicarbonate, $K_C$ is the solubility product constant for calcite, $k'_1$ is a modified forward rate constant for reaction \eqref{Ca1}, which is about $(10\sim20)k_1$.

An approximation of Fick’s first law of diffusion ~\cite{S-Z} gives the mass flux of \ce{Ca^2+} from the surface of limestone through the metal sediment layer, in our case, \ce{Fe(OH)3}, into the bulk solution
\begin{align}
J_\mathrm{d}=\frac{D}{\delta}([\ce{Ca^2+$_{\rm (s)}$}]-[\ce{Ca^2+$_{\rm (b)}$}]),
\end{align}
where $D$ (${\rm L}^2$/s) is the diffusion rate of \ce{Ca^2+} in \ce{Fe(OH)3} as porous media ~\cite{logan}, $\delta$ (L) is the thickness of the metal sediment,  \ce{[Ca^2+$_{\rm (s)}$]} is the concentration of  \ce{Ca^2+} at the surface of the limestone, and \ce{[Ca^2+$_{\rm (b)}$]} is the concentration of  \ce{Ca^2+} in the bulk solution.
For dissolution from a surface of area $A$ (${\rm m}^2$) into a fluid of volume $V$ (${\rm m}^3$), the mass balance necessitates that the rate of change of the concentration in the bulk solution is
\begin{align}\label{rateode}
\frac{d [\ce{Ca^2+}]}{d t}=\frac{A}{V} J_\mathrm{d}.
\end{align}
Combining with the rate law, we obtain
\begin{align}\label{nonlineareq}
&\frac{D}{\delta}([\ce{Ca^2+$_{\rm (s)}$}]-[\ce{Ca^2+$_{\rm (b)}$}])\nonumber\\
&\quad=\frac{1}{10}[k_1 [\ce{H+$_{\rm (s)}$}]+k_2 [\ce{H2CO3^\ast}]+k_3 [\ce{H2O}]-k_4 [\ce{Ca^2+$_{\rm (s)}$}] [\ce{HCO3-$_{\rm (s)}$}]].
\end{align}

This is overall a nonlinear equation dependent on the concentrations of calcium ions on the surface of the limestone, as well as in the water. The concentration of calcium ions in the water is obtained by solving the differential equation \eqref{rateode}. A nonlinear equation solver is then utilized to solve \eqref{nonlineareq}. 
   
Because of the low solubility of \ce{Fe^3+}, it tends to precipitate when the pH value raises to some certain threshold. To calculate the iron precipitation, we consider the following reaction and its corresponding equilibrium equation \cite{S-Z}
\begin{align}
&\ce{Fe(OH)3 + 3H+ <=> Fe^3+ + 3 H2O},\\ 
&\log[\ce{Fe^3+}]=3.2-3 \ce{pH}.
\end{align}

The effect is added in the differential equation solver to reflect the diffusion of calcium ions through the precipitaion layer on the surface of limestone. 

\section{Numerical result}\label{num}
The experiment was done by engineering students at Saint Francis University under the ``Limestone Channel at Swank 13 Mine'' project \cite{Strosnider}. The Swank 13 mine is an abandoned underground coal mine in Reade Township, northern Cambria County, Pennsylvania. An acid flow emerges into the abandoned mine and was directed down a $1000$ feet channel lined with about a foot thickness of limestone, in order to neutralize the acidity and precipitate the Fe and Al. Numerous data were collected from the field and numerical computation were conducted in Matlab to fit the real data. We used optimization toolbox to solve the nonlinear equation \eqref{nonlineareq}. $ode45$ is used to compute the change of concentrations of calcium ions along the channel. Table \ref{waterdepth}, table \ref{watervel}, and table \ref{limestone} list some sample measurement data we made for depth of water, velocity of water flow in the channel, and surface area of typical limestone at different locations in the channel, respectively.  
{\renewcommand{\arraystretch}{1.5}
\begin{table}[h]
\caption{Measurements of Water Depth}
\centering
\begin{tabular}{|c|c|c|c|c|c|c|} 
\hline
depth (m) & bottom & flag 1 & flag 2 & flag 3 & flag 4 & top \\
\hline
 {} & 0.175 & 0.320 & 0.102 & 0.101 & 0.163 & 0.310 \\ 
\hline
 {} & 0.171 & 0.264 & 0.094 & 0.114 & 0.132 & 0.274 \\  
\hline
{} & 0.182 & 0.279 & 0.124 & 0.142 & 0.119 & 0.267 \\
\hline
 {} & 0.122 & 0.292 & 0.130 & 0.132 & 0.155 & 0.271 \\
\hline
 {} & 0.111 & 0.299 & 0.140 & 0.127 & 0.172 & 0.276 \\
\hline
 average & 0.152 & 0.291 & 0.120 & 0.123 & 0.148 & 0.280 \\
\hline
\end{tabular}
\centering
\label{waterdepth}
\end{table}
\begin{table}[h]
\centering
\caption{Measurements of Water Velocity}
\begin{tabular}{|c|c|c|c|} 
\hline
{}& velocity (m/s) & error of velocity (m/s) & angle of velocity (m/$\text{s}^2$) \\
\hline
bottom & 0.303 & 0.025 & -11 \\ 
\hline
flag 1 & 0.238 & 0.007 & 15 \\  
\hline
flag 2 & 0.572 & 0.011 & -6 \\
\hline
flag 3 & 0.682 & 0.028 & -14 \\
\hline
flag 4 & 0.587 & 0.021 & 3 \\
\hline
top & 0.299 & 0.291 & 0 \\
\hline
average & 0.477 & 0.016 & 2 \\
\hline
\end{tabular}
\centering
\label{watervel}
\end{table}
\begin{table}
\centering
\caption{Measurements of Limestone}
\begin{tabular}{|c|c|c|c|} 
\hline
surface area ($\text{m}^2$) & height (m) & width (m) & length (m) \\
\hline
0.047 & 0.127 & 0.089 & 0.056 \\ 
\hline
0.027 & 0.102 & 0.058 & 0.046 \\  
\hline
0.088 & 0.206 & 0.107 & 0.071 \\
\hline
0.033 & 0.058 & 0.066 & 0.043 \\
\hline
0.015 & 0.094 & 0.091 & 0.043 \\
\hline
\end{tabular}
\label{limestone}
\end{table}
The numerical result obtained by computation are shown in figure \ref{figure1} and figure \ref{figure2}.   
\begin{figure}[h]
\begin{center}
\resizebox*{12cm}{!}{
\includegraphics{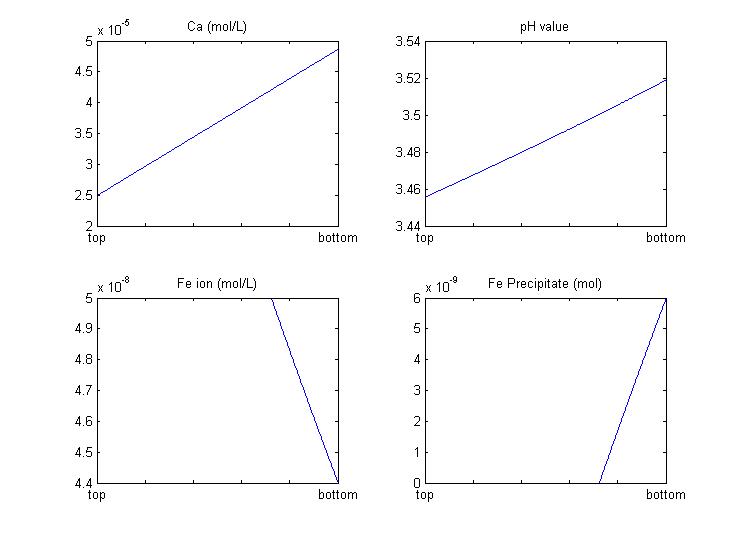}}
\caption{Change of Concentration of Different Species } \label{figure1}
\end{center}
\end{figure}
\begin{figure}[h]
\begin{center}
\resizebox*{12cm}{!}{
\includegraphics{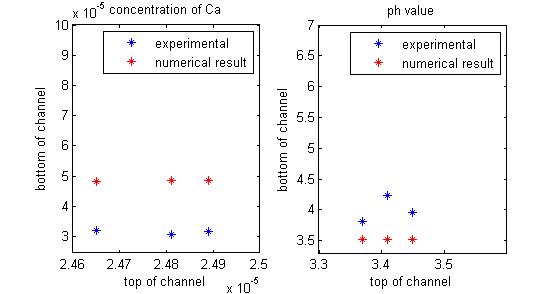}}
\caption{Comparison of Results} \label{figure2}
\end{center}
\end{figure}

Figure \ref{figure1} shows the change of concentration of calcium and iron ions along the channel, as well as the change of pH value. We can observe the release of calcium and pH value is increased accordingly. When the pH value achieves a threshold value, iron precipitates start to appear. Figure \ref{figure2} shows the comparison of our numerical results and measurements from the channel with three sets of data. Both show the predicted improvement of water quality.

\section{conclusion and future discussion} \label{concl}
In this project we have set up a mathematical model of open limestone channel using Fick's first law and the rate law of calcite dissolution. The diffusion process is approximated by linear decrease, which is the steady state equation. The computational results suggest that without diffusion, the acid water could be completely neutralized. However, both numerical and experimental data show that the effect of open limestone channel treatment is greatly reduced by the iron sediments, which is consistent with the conclusion made in ~\cite{Z-S-R}. There are several other factors to be considered in our model in the future. The velocity of water flow in the channel may take away some of the iron sediments. A coefficient may be added accordingly. A time delay may be considered in the diffusion process.  

\section*{Acknowledgement}
The authors would like to thank Dr. Bill Strosnider and his student Amanda Conrad for their generous help on the numerous field trips to the Swank 13 mine, and for providing the chemical data and project materials.

\section*{Funding}
This work was supported by the National Science Foundation under Grant No. DUE-1161227.

\end{document}